\documentclass{jfm}

\usepackage{afterpage}
\usepackage{longtable}
\usepackage{hyperref}
\usepackage[english]{babel}
\usepackage{dcolumn}
\usepackage{bm}
\usepackage{pifont}
\usepackage{amsfonts}
\usepackage{amsmath}
\usepackage{upmath}
\usepackage{amssymb}
\usepackage{wasysym}
\usepackage{amsbsy}
\usepackage{graphicx}
\usepackage{natbib}
\usepackage{psfrag}
\usepackage{lineno}
\usepackage{color}
\usepackage{multirow}
\usepackage{cases}
\usepackage{stackengine}
\usepackage{float}
\usepackage{blkarray}
\usepackage{cancel}
\usepackage{mathrsfs}
\usepackage{empheq}

\newcommand{\D}{\ensuremath{\mathrm{d}}}
\providecommand\bnabla{\boldsymbol{\nabla}}
\providecommand\bcdot{\boldsymbol{\cdot}}
\providecommand\upi{\pi}%

\newcommand{\V}{\mathcal{V}}
\newcommand{\Surf}{\partial \V}

\newcommand{\Oh}{\mbox{\textit{Oh}}} 	        % Ohnesorge number
\newcommand{\Ut}{\skew3\tilde{U}}
\newcommand{\Vt}{\skew3\tilde{V}}

\newcommand{\Pt}{\skew3\tilde{P}}
\newcommand{\hht}{\skew3\tilde{h}}

\newcommand{\hmin}{h_\textrm{min}}
\newcommand{\hmind}{h_\textrm{min}^*}
\newcommand{\tauxy}{\tau_{xy}}
\newcommand{\umax}{u_\textrm{max}}
\newcommand{\tr}{t_\textrm{R}}

\title[Inertial rupture of ultrathin liquid films]%
{Inertial rupture of ultrathin liquid films}

\author%
[D. Moreno-Boza, A. Mart\'inez-Calvo, A. Sevilla]
{D. Moreno-Boza\thanks{Email address for correspondence: damoreno@pa.uc3m.es},
\ns A. Mart\'inez-Calvo,\ns and A. Sevilla}

\affiliation{%
Grupo de Mec\'anica de Fluidos,
Departamento de Ingenier\'ia T\'ermica y de Fluidos,
Universidad Carlos III de Madrid,
Av.~Universidad 30, 28911 Legan\'es (Madrid),
Spain\\[\affilskip]}

\begin{document}
\maketitle
%\linenumbers

\begin{abstract}
Theory and numerical simulations of the Navier-Stokes equations are used to unravel the inertia-driven dewetting dynamics of an ultrathin film of Newtonian liquid deposited on a solid substrate. A classification of the film thinning regimes at finite Ohnersorge numbers is provided, unifying previous findings. We reveal that, for Ohnesorge numbers smaller than one, the final approach to the rupture singularity close to the molecular scales is controlled by a balance between liquid inertia and van der Waals forces leading to a self-similar asymptotic regime where $\hmin \propto \tau^{2/5}$ as $\tau \to 0$, where $\hmin$ is the minimum film thickness and $\tau$ is the time remaining before rupture. The flow exhibits a three-region structure comprising an irrotational core delimited by a pair of boundary layers at the wall and at the free surface. A potential-flow description of the irrotational core is provided, which is asymptotically matched with the viscous layers, allowing us to present a complete parameter-free asymptotic description of inertia-driven film rupture.
\end{abstract}

\begin{keywords}
Thin film, Instability, Dewetting
\end{keywords}

\section{Introduction}
\label{sec:intro}

A growing number of emergent technologies involve the manipulation of liquid metals from millimetric to sub-micron scales~\citep{Dickey2014,debroy2018additive,kondic2019liquid}. For instance, in additive manufacturing and electronics, liquid gallium has received special attention, mainly due to its non-toxic character and its low melting point, just slightly above room temperature. In addition, plasmonic devices and many patterning and coating processes, rely on metal or metal-like materials like silver and gold. Many of these applications involve the presence of liquid jets, liquid droplets and thin liquid films in critical intermediate stages. In particular, thin liquid films have been extensively studied due to their central role in many engineering devices, as well as in geological and physiological flows, to cite a few~\citep[the reader is referred to][for thorough and excellent reviews]{de1985wetting,Oron1997,Bonn2009,craster2009dynamics}. In most previous theoretical studies of liquid film dynamics the effect of liquid inertia has been neglected, an approximation that is highly accurate when the characteristic length scale is small and the liquid viscosity is large. In contrast, inertia cannot be neglected when the working fluid is a liquefied metal, as clearly pointed out in a number of recent investigations~\citep{Gonzalez2013,Krishna2009,Mckeown2012,Gonzalez2016,kondic2019liquid}.

In the present work, we report a theoretical and numerical study of the influence of liquid inertia on the instability, nonlinear dynamics and rupture of ultra-thin liquid films placed on a solid impermeable substrate, which are known to become unstable in the non-wetting case due to the action of the long-range van der Waals (vdW) forces. In particular, we unravel the self-similar regimes that are transiently achieved as contact is approached, allowing us to provide a unified description of all previous theoretical findings. We also deduce the correct self-similar regime that leads to the contact singularity, in which the minimum film thickness scales with the time to contact as $\hmin\propto\tau^{2/5}$, stemming from a balance between liquid inertia and vdW forces, the capillary force being asymptotically subdominant. The latter result is in marked contrast with the prevailing descriptions of the inertial limit, which predict $\hmin\propto\tau^{2/7}$ from a balance of liquid inertia with capillary and vdW forces~\citep{ZhangLister1999,garg2017self}.

The paper is organised as follows. In \S2 we present the mathematical model used to describe the inertial dewetting flow, and we present numerical integrations of the Navier-Stokes equations aimed at establishing its main dynamical features. In \S3 we first provide an exhaustive classification of the asymptotic self-similar regimes during film thinning at finite Ohnesorge numbers. We then focus on the inertial near-rupture flow, including the universal self-similar potential flow in the bulk of the liquid film, as well as the wall and free-surface boundary layers. Some concluding remarks are finally presented in \S4.

\section{Governing equations and numerical results}

\subsection{Navier-Stokes equations}

We consider the incompressible flow resulting from the thinning of a dewetting two-dimensional thin liquid film of density $\rho$, viscosity $\mu$, and initial thickness $h^{*}_o$, initially resting on a solid substrate and separated from a passive immiscible ambient by an interface of constant surface tension coefficient $\sigma$. Such films are known to become unstable to infinitesimal perturbations due to long-range intermolecular forces when the film thickness lies below a thickness threshold of about the 100~$\textrm{nm}$~\citep{Vrij1966,Blossey2012}. To describe the resulting unstable dynamics, we make use of the incompressible Navier-Stokes equations, which are non-dimensionalised upon taking
\begin{equation}\label{eq:scales}
\ell_c = a = \sqrt{\frac{A}{6\upi \sigma}}, \quad v_c = \sqrt{  \frac{A}{6 \upi \rho a^3} }, \quad t_c = \sqrt{\frac{6 \upi \rho a^5}{A}}, \quad p_c = \phi_c = \frac{A}{6 \upi a^3},
\end{equation}
as the characteristic length, velocity, time, pressure, and intermolecular potential scales, respectively, where $a$ is the molecular length scale, to yield
\begin{subequations}\label{eq:ns}
\begin{alignat}{2}
&\bnabla \bcdot \bm{v} = 0, \quad \bm{x} \in \V \label{eq:continuity}\\
&\partial_t \bm{v} + \bm{v} \bcdot \bnabla \bm{v} = -\bnabla\left( p + h^{-3} \right) + \Oh \bnabla^2 \bm{v}, \quad \bm{x} \in \V \label{eq:momentum}
\end{alignat}
\end{subequations}
% \begin{subequations}\label{eq:ns}
% \begin{gather}
% \bnabla \bcdot \bm{v} = 0, \quad  \text{and} \quad \partial_t \bm{v} + \bm{v} \bcdot \bnabla \bm{v} = -\bnabla\left( p + h^{-3} \right) + \Oh \bnabla^2 \bm{v}, \quad \bm{x} \in \V, \tag{\theequation $a$,$b$} 
% \end{gather}
% \end{subequations}
as the relevant equations of motion, where $\Oh = \mu/\sqrt{\rho a \sigma}=\mu[6\upi/(\rho^2 A \sigma)]^{1/4}$ is the Ohnesorge number based on $a$, $\V$ is the liquid film domain, $\bm{x} = (x,y)$, and $\bm{v} = (u,v)$ is the velocity field, assumed to be two-dimensional. Note that we have employed the simplest intermolecular potential, namely $\phi = h^{-3}$, derived from the case of the vdW force between two parallel surfaces. At the free surface $\Surf$ we impose the kinematic and stress balance boundary conditions, which read
\begin{subequations}\label{eq:interface}
\begin{alignat}{2}
&\bm{n}\bcdot\left( \partial_t \bm{x}_s - \bm{v} \right) = 0, \quad \bm{x} \in \Surf \label{eq:interface_kinematic} \\
&-p\bm{n}+\Oh \left(\bnabla{\bm{v}}+\bnabla{\bm{v}}^\mathrm{T}\right)\bcdot \bm{n} = - \bm{n} (\bnabla \bcdot \bm{n}),  \quad \bm{x} \in \Surf \label{eq:interface_stress}
\end{alignat}
\end{subequations} 
% \begin{equation}\label{eq:interface}
% \bm{n}\bcdot\left( \partial_t \bm{x}_s - \bm{v} \right) = 0,  \quad  \text{and} \quad -p\bm{n}+\Oh \left(\bnabla{\bm{v}}+\bnabla{\bm{v}}^\mathrm{T}\right)\bcdot \bm{n} = - \bm{n} (\bnabla \bcdot \bm{n}),  \quad \bm{x} \in \Surf \tag{2.3$a$,$b$}
% \end{equation} 
respectively, where $\bm{x}_s$ is the parameterisation of the interface, located at $y = h(x,t)$, and $\bm{n}$ is the unit normal vector to the interface. At the solid substrate, $y = 0$, the no-slip boundary condition is enforced, $\bm{v} = \bm{0}$. As for the initial conditions, in the numerical simulations we consider half wavelength of a spatially periodic liquid film, and thus we impose the symmetry condition $u = 0$ at $x = 0$ and $x = \upi/k$, where $k < k_c = \sqrt{3}/h_o^2$ is the dimensionless wavenumber of the initially perturbed interface $\bm{x}_s = \left[x, h_o \, (1-\epsilon \cos{kx})\right]$, imposed at $t = 0$. Here, $k_c$ is the dimensionless cut-off wavenumber predicted by linear instability theory~\citep{Vrij1966}, $h_o = h^{*}_o/a$ is the initial film thickness normalised with the molecular length scale, and $\epsilon$ is a small positive constant that triggers the instability and induces the rupture of the liquid film at $x=0$ and $t=\tr$. Finally, note that two non-dimensional parameters govern the flow at hand, namely the dimensionless initial film thickness, $h_o$, and the Ohnesorge number, $\Oh$.

%Two non-dimensional parameters govern the flow at hand, namely the dimensionless initial film thickness, $h_o$, and the molecular Ohnesorge number, $\Oh$. These two parameters can be written in terms of the \emph{global} Weber and Ohnesorge numbers, $\Web_o$ and $\Oh_o$, defined using $h_o^*$ instead of $a$ as the characteristic length scale, and a corresponding characteristic velocity $v_{co}$ given by the balance $\rho v_{co}^2/h_o^* \sim A/(6\pi h_o^{*4})\Rightarrow v_{co}^2=A/(6\pi\rho\sigma h_o^{*3})$, to be compared with the the velocity $v_c$ appearing in~\eqref{eq:scales}, which uses $a$ as characteristic length scale. Using these global scales, the corresponding Weber and Ohnesorge numbers are $\Web_o=\rho v_{co}^2 h_o^*/\sigma=h_o^{-2}$ and $\Oh_o=\mu/\sqrt{\rho\sigma h_o^*}=\Oh/\sqrt{h_o}=\Oh\Web_o^{1/4}$.

\subsection{Numerical simulations of the flow evolution towards the singularity}

\begin{figure}
    \centering
    \includegraphics[width=1.0\textwidth]{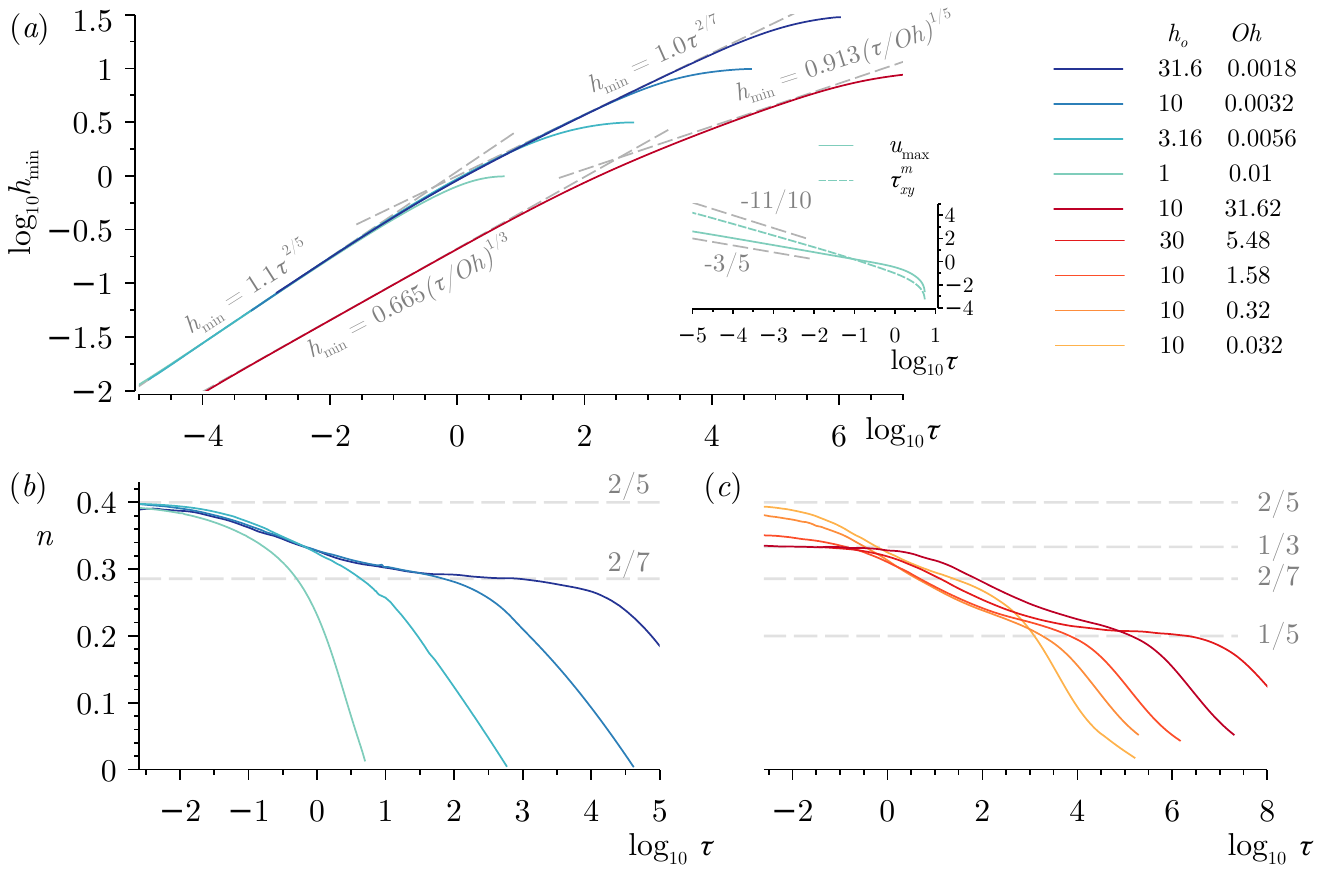}
    \caption{($a$) The function $\log_{10}\hmin(\log_{10}\tau)$ for different values of $h_o$ and $\Oh$ indicated in the legend. The gray dashed lines show the self-similar thinning laws associated to regimes I--IV in table~\ref{tab:table1}. The inset displays the maximum streamwise velocity, $\umax$, and the maximum shear stress at the wall $\tauxy^m$, for $h_o = 1$ and $\Oh = 10^{-2}$. The instantaneous exponent $n(\tau)$: ($b$) influence of $h_o$ for $\Oh\ll 1$, and ($c$) influence of $\Oh$ for $h_o = (10,30)$.}
    \label{fig:fig1}
\end{figure}

The finite-element method recently employed by~\cite{morenobozaetal2020} was slightly modified to integrate~\eqref{eq:ns}--\eqref{eq:interface} for a wide range of values of $\Oh$ and $h_o$. As the film evolved towards rupture, we tracked the minimum film thickness $\hmin = h(0,t)$, the maximum streamwise velocity, $\umax$, and the maximum wall shear stress, $\tauxy = \left. \partial_y u \right|_{y=0}$. Note that the existence of a power law $\hmin\propto\tau^n$ for some value of $n > 0$, where $\tau = \tr-t$ is the time remaining to rupture, is a local self-similarity test that can be easily extracted from the numerical simulations. Figure~\ref{fig:fig1} shows several representative film evolutions revealing that $(\hmin,\umax,\tauxy)\propto (\tau^{2/5},\tau^{-3/5},\tau^{-11/10})$ as $\tau\to 0$, for all the values of $h_o$ and sufficiently small values of $\Oh$. However, it is important to emphasise that the latter universal behaviour is achieved for $(\tau,\hmin)\lesssim (0.01,0.16)$, for which the continuum approximation is compromised. We thus conclude that the $2/5$ power law would never be fully established under realistic conditions. For $\tau\gtrsim 0.01$ and $h_o\lesssim 30$, the results of figure~\ref{fig:fig1} show no sign of sustained power-law behaviour for $\Oh\lesssim 1$, as revealed by the instantaneous exponents $n(\tau)=\D\log_{10}\hmin/\D\log_{10}\tau$ plotted in figures~\ref{fig:fig1}($b$) and~\ref{fig:fig1}($c$).

In particular, figure~\ref{fig:fig1}($b$) reveals that, for all the cases where $\Oh\ll 1$, the value of $n$ increases monotonically with $n(\tau)\to 2/5$ as $\tau\to 0$ independently of $h_o$. It is also deduced that the function $n(\tau)$ has no inflection points for the smallest values of $h_o$, namely $h_o=1$ and $h_o=3.16$, indicating the absence of intermediate self-similarity other than the $2/5$ power-law which, as shown in~\S\ref{subsec:oom}, is due to a balance between liquid inertia and vdW forces with negligible surface tension and viscous forces (regime IV of table~\ref{tab:table1}). However, for $h_o=10$ and $h_o=31.6$, an inflection point appears in the evolution of $n(\tau)$ near the value $n=2/7$, which corresponds to the regime described by~\citet{garg2017self} as a local balance between liquid inertia, vdW forces and surface tension forces (regime III of table~\ref{tab:table1}). Although for $h_o\leq 31.6$ the $2/7$-regime is only established for less than a decade, its range of validity increases with increasing $h_o$. Indeed,~\citet{garg2017self} reported Navier--Stokes simulations for $h_o=1165$, in which case the $2/7$ regime is established for several decades for a shear-thinning liquid.
%However, it is important to note that the latter value of $h_o$ is unrealistically large, in that it leads to a rupture time $\tr\sim 2\times 10^{11}$ during which other effects, like gravitational or capillary drainage, would affect the film evolution.

The effect of liquid viscosity on the rupture regimes is deduced from figure~\ref{fig:fig1}($c$), where $n(\tau)$ is plotted for several values of $\Oh$ and $h_o=(10,30)$. As the value of $\Oh$ increases, the film is seen to pass through a number of intermediate self-similar regimes described in previous studies, although none of them are clearly established except in the limit $\tau\to 0$. As $\tau$ decreases, the $1/5$ power-law described by~\citet{ZhangLister1999} (regime I of table~\ref{tab:table1}) is first observed during a brief transient, followed by the $2/7$-law for small enough values of $\Oh$. For smaller values of $\tau$, the $1/3$-law described by~\citet{morenobozaetal2020} (regime II in table~\ref{tab:table1}) is clearly observed, holding until the singularity for $\Oh=31.62$. Finally, regime IV is reached as $\tau\to 0$ for $\Oh\lesssim 1$.

As shown in figure~\ref{fig:fig2}($a$) for $\Oh = 10^{-2}$ and $h_o = 1$, the local rupture flow presents a distinctive multi-scale structure. The vorticity concentrates in boundary layers at the wall and at the free surface, surrounding a central irrotational core. The self-similar potential flow and the relevant scalings of the boundary layers are described in~\ref{subsec:2over5}.

\section{The self-similar regimes}
\label{sec:regimes}

\subsection{The thinning regimes at finite Ohnesorge numbers: order-of-magnitude analysis}
\label{subsec:oom}

The results of figure~\ref{fig:fig1} reveal the existence of different regimes during the thinning of the liquid film for finite values of $\Oh$, which will now be explained using order-of-magnitude estimations of the governing equations~\eqref{eq:ns}-\eqref{eq:interface}. Hereinafter, the dimensional variables with a non-dimensional counterpart will be denoted with an asterisk. We begin by defining two key variables, namely the local slenderness, $\epsilon_l(\tau_*)=\hmind/x_c^*$, and the local reduced Reynolds number, $\epsilon_l\Rey_l$, where $\Rey_l=\rho u_c^* \hmind/\mu$. Here, $x_c^*(\tau_*)$ and $u_c^*(\tau_*)$ are the characteristic longitudinal length scale and velocity, respectively, while the local curvature is $\kappa^*\sim \hmind/x_c^{*2}$, such that $\kappa^*\hmind\sim \epsilon_l^2$. The continuity equation~\eqref{eq:continuity} provides the estimate $v_c^*\sim u_c^*\epsilon_l$, where $v_c^*\sim \hmind/\tau_*$ is the characteristic transverse velocity. Thus, $\epsilon_l\Rey_l=\rho v_c^*\hmind/\mu\sim \rho \hmin^{*2}/(\mu\tau_*)$.

The momentum equation~\eqref{eq:momentum} admits two limiting dominant balances depending on the value of $\epsilon_l\Rey_l$. The viscous regime prevails when $\epsilon_l\Rey_l\ll 1$, leading to the vdW-viscous balance $A/(\hmin^{*3}\,x_c^*) \sim \mu u_c^*/\hmin^{*2} \sim \mu v_c^* x_c^*/\hmin^{*3} \Rightarrow x_c^* \sim (A/\mu)^{1/2}\tau_*/\hmind$ and $\epsilon_l^2 \sim (\mu/A)\,\hmin^{*3}/\tau_*$. Assuming $\hmind \propto \tau_*^{\alpha}$, one has $\epsilon_l^2 \propto \tau_*^{3\alpha-1}$, so that asymptotic slenderness requires that $\alpha>1/3$. Two different values of $\alpha$ are possible depending on $\epsilon_l$. When $\epsilon_l\ll 1$, the capillary pressure gradient is $\sigma \partial_{x^*} \kappa^*\sim \sigma \hmind/x_c^{*3}$, and its relative importance over the driving vdW force is $\sim (\sigma/A)\hmin^{*4}/x_c^{*2} \sim (\mu\sigma/A^2)\hmin^{*5}/\tau_* \propto \tau_*^{5\alpha-1}$, whence $\alpha\geq 1/5$. When $\alpha>1/5$, surface tension forces are asymptotically negligible, while vdW, viscous, and surface tension forces are in balance when $\alpha=1/5$. The latter scenario, which holds when $\epsilon_l\ll 1$, leads to the rupture law $\hmind\sim [A^2/(\mu\sigma)]^{1/5}\tau_*^{1/5}$~\citep{ZhangLister1999}, with $\epsilon_l^2\sim \kappa^*\hmind \sim (\mu^2 A/\sigma^3)^{1/5}\tau_*^{-2/5}$ and $\epsilon_l\Rey_l \sim [\rho A^{4/5}/(\mu^{7/5}\sigma^{2/5})]\tau_*^{-3/5}$. The latter solution fails either when $\epsilon\sim 1$ due to the breakdown of slenderness, or when $\epsilon_l\Rey_l \sim 1$ due to inertial effects. These conditions hold at the crossover times $\tau_*=\tau_{*\epsilon}\sim \mu A^{1/2}/\sigma^{3/2}$ and $\tau_*=\tau_{*\rho} \sim \rho^{5/3}A^{4/3}/(\mu^{7/3}\sigma^{2/3})$, respectively, with associated minimum thicknesses $\hmind(\tau_{*\epsilon})\sim (A/\sigma)^{1/2} \sim a$ and $\hmind(\tau_{*\rho})\sim \rho^{1/3}A^{2/3}/(\mu^{2/3}\sigma^{1/3})$ which, in dimensionless terms, read $\hmin(\tau_{\epsilon})\sim 1$ and $\hmin(\tau_{\rho})\sim \Oh^{-2/3}$. The former condition implies a breakdown of the $1/5$ power law close to the molecular scale~\citep{morenobozaetal2020}, while the latter one marks an inertial crossover at a length scale that depends on $\Oh$, with two relevant limiting cases. When $\hmin(\tau_{\rho}) \gtrsim h_o \Rightarrow \Oh \lesssim h_o^{-3/2}$, liquid inertia is important from the onset film thinning, in which case the $1/5$ power law is never established. On the other hand, when $\hmin(\tau_{\rho})\lesssim 1\Rightarrow \Oh \gtrsim 1$, liquid inertia becomes important at molecular scales, and the $1/5$-law holds until $\hmin \sim 1$, when slenderness breaks down. In the intermediate case $h_o^{-3/2}\lesssim \Oh \lesssim 1$, the $1/5$ law experiences the inertial crossover at $\hmin \sim \Oh^{-2/3}$. In the case $\Oh\gtrsim 1$, the breakdown of slenderness that occurs when $\epsilon_l \sim 1$ gives rise to a regime given by the balance $A/(\hmin^{*4})\sim \mu u_c^*/\hmin^{*2} \sim \mu /(\hmind\tau_*) \Rightarrow \hmind\sim(A/\mu)^{1/3}\tau_*^{1/3}$~\citep{morenobozaetal2020}.

The inertial regime established when $\epsilon_l\Rey_l\gg 1$ depends on the local slenderness. We consider first the case with $\epsilon_l\ll 1$, for which the balance of liquid inertia and vdW forces yields $\rho u_c^{*2} \sim A/\hmin^{*3} \Rightarrow \hmin^{*3}x_c^{*2} \sim (A/\rho)\tau_*^{2}$. The ratio between the capillary pressure gradient, $\sigma\hmind/x_c^{*3}$, and the inertial force, is $(\rho\sigma/A^2)\hmin^{*7}\tau_*^{-2} \sim (\rho\sigma/A^2)\tau_*^{7\alpha-2}$, whence $\alpha\geq 2/7$. The case $\alpha=2/7$ corresponds to a balance of liquid inertia with vdW and surface tension forces~\citep{garg2017self}, and leads to the thinning law $\hmind \sim [A^{2}/(\rho\sigma)]^{1/7}\tau_*^{2/7}$. During this regime, $\epsilon_l^2\sim \kappa^*\hmind \sim (\rho^{2/7}A^{3/7}/\sigma^{5/7})\tau_*^{-4/7}$, so that slenderness breaks down when $\tau_*=\tau_{*\epsilon}\sim \rho^{1/2}A^{3/4}/\sigma^{5/4}\Rightarrow \hmind(\tau_{*\epsilon}) \sim (A/\sigma)^{1/2}\sim a$. Thus, the slender inertial regime crosses over to a non-slender inertial regime when $\hmin \sim 1$, in which the inertia-vdW balance writes $\rho \hmin^{*2}/\tau_*^2 \sim A/\hmin^{*3}$, so that $\hmind \sim (A/\rho)^{1/5}\tau_*^{2/5}$, and the relative importance of the capillary pressure gradient writes $[\sigma^{10/7}/(\rho^{4/7}A^{6/7})]\tau_*^{8/7}\to 0$ as $\tau_*\to 0$, so that the ultimate regime reached just prior to rupture is a consistent dominant balance between inertia and vdW forces, with asymptotically negligible surface tension forces.

The previous development demonstrates the existence of four regimes during the vdW-induced rupture of the film at finite values of $\Oh$, classified in table~\ref{tab:table1}. In dimensionless terms, these regimes are: I) When $\epsilon_l\Rey_l\ll 1$ and $\epsilon_l\ll 1$, the slender viscous regime discovered by~\citet{ZhangLister1999} holds, with a minimum thickness $\hmin \sim (\tau/\Oh)^{1/5}$ and a local curvature $\kappa \sim (\tau/\Oh)^{-3/5}$. II) When $\epsilon_l\Rey_l\ll 1$ and $\epsilon_l \sim 1$, the non-slender Stokes flow described by~\citet{morenobozaetal2020} takes place, with $\hmin \sim (\tau/\Oh)^{1/3}$ and $\kappa \sim 1$. III) When $\epsilon_l\Rey_l\gg 1$ and $\epsilon_l\ll 1$, the slender inertial regime reported by~\citet{garg2017self} prevails, with $\hmin \sim \tau^{2/7}$ and $\kappa \sim \tau^{-6/7}$. IV) Finally, when $\epsilon_l\Rey_l\gg 1$ and $\epsilon_l\sim 1$ an ultimate non-slender inertial regime takes over, with $\hmin \sim \tau^{2/5}$ and $\kappa \sim 1$.

%%%%%%%%%%%%%%%%%%%%%%%%%%%
\begin{table}
\centering
\begin{tabular}{c c c c c c c}
Regime & Validity & Balance & Realisability & $\hmin(\tau)$ & $\kappa(\tau)$ & Reference \\
I & $\epsilon_l\Rey_l \ll 1$, $\epsilon_l\ll 1$ & vdW-$\mu$-$\sigma$ & $\Oh\gtrsim h_o^{-3/2}$ & $(\tau/\Oh)^{1/5}$ & $(\tau/\Oh)^{-3/5}$ &  ZL99 \\
II & $\epsilon_l\Rey_l \ll 1$, $\epsilon_l \sim 1$ & vdW-$\mu$ & $\Oh\gtrsim h_o^{-3/2}$ & $(\tau/\Oh)^{1/3}$ & $\sim 1$ &  MB20 \\
III & $\epsilon_l\Rey_l \gg 1$, $\epsilon_l\ll 1$ & vdW-$\rho$-$\sigma$ & $\Oh\lesssim 1$ & $\tau^{2/7}$ & $\tau^{-6/7}$ &  GA17\\
IV & $\epsilon_l\Rey_l \gg 1$, $\epsilon_l \sim 1$ & vdW-$\rho$ & $\Oh\lesssim 1$ & $\tau^{2/5}$ & $\sim 1$ & Present work
\end{tabular}
\caption{The four thinning regimes during the rupture of ultrathin liquid films. In the balances, $\rho$ and $\mu$ stand for liquid inertia and viscous forces, respectively. The references stand for~\citet{ZhangLister1999} (ZL99),~\citet{morenobozaetal2020} (MB20) and~\citet{garg2017self} (GA17).\label{tab:table1}}
\end{table}
%%%%%%%%%%%%%%%%%%%%%%%%%%%

\subsection{The universal self-similar regime near rupture}
\label{subsec:2over5}

Suggested by the flow evolution shown in figure~\ref{fig:fig1}, and by the order-of-magnitude analysis presented in~\S\ref{subsec:oom}, the universal behaviour $\hmin = 1.1\,\tau^{2/5}$ associated to regime IV in table~\ref{tab:table1}, will now be unveiled using similarity theory.

\subsubsection{The self-similar potential flow}
\label{subsubsec:self}

Letting $x = \tau^{\beta} \xi$, $y = \tau^{\alpha} \eta$, $u = \tau^\gamma U$, $v = \tau^{\gamma+\alpha-\beta}V$, $p = \tau^{-3\alpha} P$, $h = \tau^\alpha f(\xi)$, where $\alpha$, $\beta$, and $\gamma$ are real numbers, a consistent leading-order dominant balance between inertia and vdW forces can indeed be found by substituting the similarity test into~\eqref{eq:ns} and~\eqref{eq:interface}, and performing the limit $\tau\to 0$, yielding the exponents $\alpha=\beta=2/5$ and $\gamma=\beta-1=-3/5$, in agreement with the results of~\S\ref{subsec:oom}. To unravel the structure of the leading-order potential flow as $\tau\to 0$ we make use of the velocity potential $\Phi$ such that $U = \Phi_\xi$ and $V = \Phi_\eta$, reducing the description to the integration of Laplace equation,
\begin{equation}
    \label{eq:PHI}
    \Phi_{\xi\xi} + \Phi_{\eta\eta} =0,  
\end{equation} 
in $0 < \xi < \infty$ and $0 < \eta < f(\xi)$, where $f(\xi)$ is the {\em a-priori} unknown shape of the free surface. Hereinafter, subscripts denote partial derivatives unless stated otherwise. The leading-order contribution of the stress balance at the interface reduces to an Euler-Bernoulli-like boundary condition, to be imposed together with the kinematic condition:
\begin{subequations}\label{ssBCs}
\begin{alignat}{2}
    & \frac{1}{f^3} + \frac{1}{5}\Phi + \frac{2}{5} \left(\xi \Phi_\xi + \eta \Phi_\eta \right) + \frac{1}{2}\left(\Phi_\xi^2+\Phi_\eta^2\right) = 0,  \label{eq:ssEulerBernoulli} \\
    & \frac{2}{5} f + \Phi_\eta - f_\xi \left( \frac{2}{5}\xi + \Phi_\xi \right) = 0, \label{eq:ssKinem}
\end{alignat} 
\end{subequations}
along $\eta = f$, together with the no-penetration condition $\bm{n} \bcdot \bnabla \Phi = 0$ at $\eta=0$, where the gradient operator is applied with respect to $\xi$ and $\eta$. Problems of similar mathematical nature were derived for instance by~\cite{zeff2000singularity} in the context of jets emerging from Faraday waves, and by~\cite{morenobozaetal2020} to describe the self-similar Stokes flow leading to thin-film rupture for $\Oh\gtrsim 1$. Equations~\eqref{eq:ns}--\eqref{eq:ssKinem} were integrated using an algorithm similar to that described in~\cite{morenobozaetal2020}, providing the free surface as part of the solution.

\begin{figure}
    \centering
    \includegraphics[width=\textwidth]{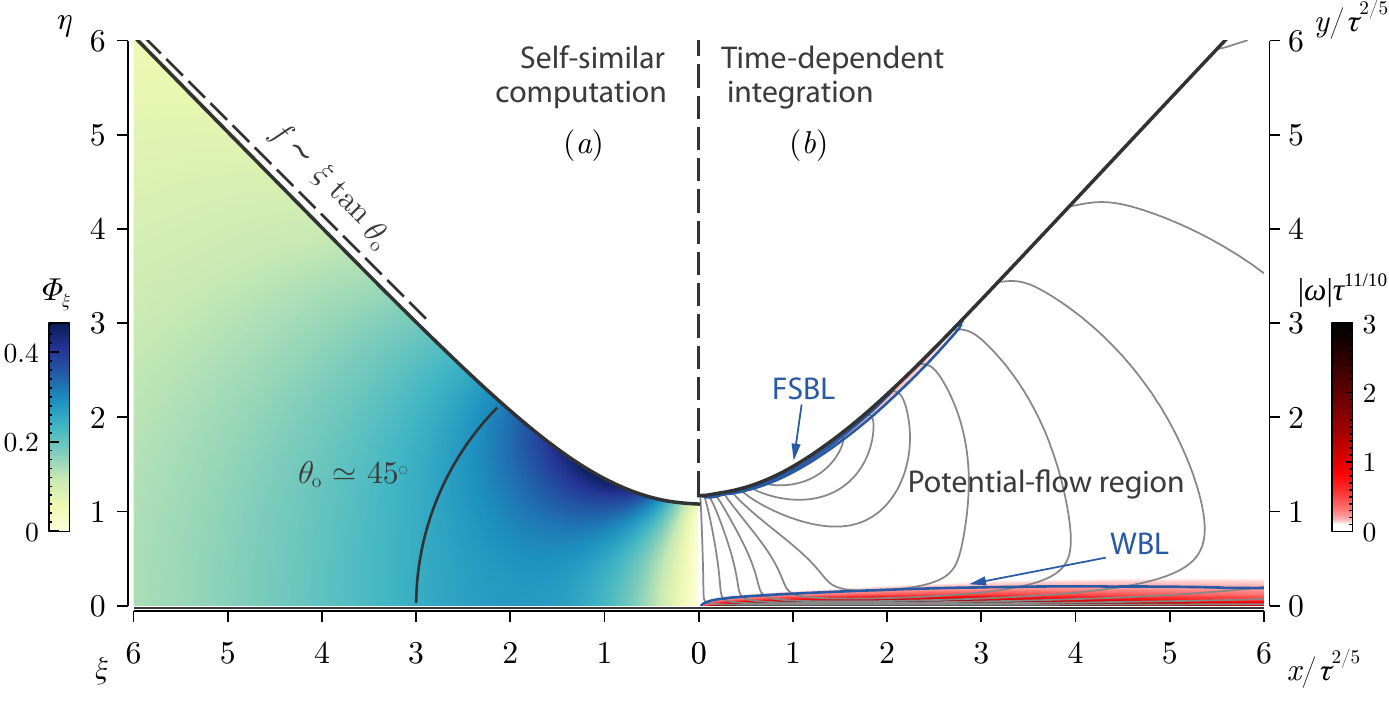}
    \caption{(a) Contours of longitudinal velocity, $\Phi_{\xi}(\xi,\eta)$, obtained from the self-similar potential flow. The free surface is wedge-shaped, with an angle $\theta_o=45^{\circ}$ off the solid wall. (b) Snapshot of a numerical simulation performed for $\Oh=0.01$ and $h_o=1$ at $\tau=4.41\times 10^{-6}$, with $\epsilon = 0.01$ and $k=0.3k_c$. Solid lines represent contours of longitudinal velocity, and colours represent contours of normalised vorticity, $|\omega|\tau^{11/10}$.}
    \label{fig:fig2}
\end{figure}

The solution, shown in figure~\ref{fig:fig2}, exhibits a wedge-shaped film $f\sim (\xi-\xi_o) \tan \theta_o$, for some value of $\xi_o$, as $r \to \infty$ characterised by an opening angle $\theta_o \simeq 45^\circ$ off the substrate, where $r^2 = (\xi-\xi_o)^2 + \eta^2$ and $\theta = \arctan{\eta/\xi}$ are polar coordinates. Examination of the far-field revealed a consistent radial decay of the potential of the form $\Phi \sim r^{-\lambda}$ along the ray $\theta = \theta_o$. The value of $\lambda$ was seen to adjust quite well to $1/2$ from the numerical computation (see inset of figure~\ref{fig:fig3}a). This is indeed the only value of $\lambda$ that ensures compatibility in terms of powers of $r$ in~\eqref{eq:ssEulerBernoulli}. A noteworthy aspect of the result shown in figure~\ref{fig:fig2} is the displacement effect of the wall boundary layer on the outer irrotational core. Indeed, the Navier-Stokes solution of figure~\ref{fig:fig2}($b$) is seen to be slightly displaced in the positive $y$-direction with respect to the potential flow of figure~\ref{fig:fig2}($a$).

The non-slender universal regime described in the present section is the inertial counterpart of the self-similar Stokes flow reported by~\citet{morenobozaetal2020}, respectively classified as regimes IV and II in table~\ref{tab:table1}. It is noteworthy that the local interface shape in the Stokes case is also a wedge, but with an opening angle of $37^{\circ}$, instead of the larger angle of $45^{\circ}$ associated with inertial break-up.

\subsubsection{The boundary layers}
\label{subsubsec:BL}

\begin{figure}
    \centering
    \includegraphics[width=392pt]{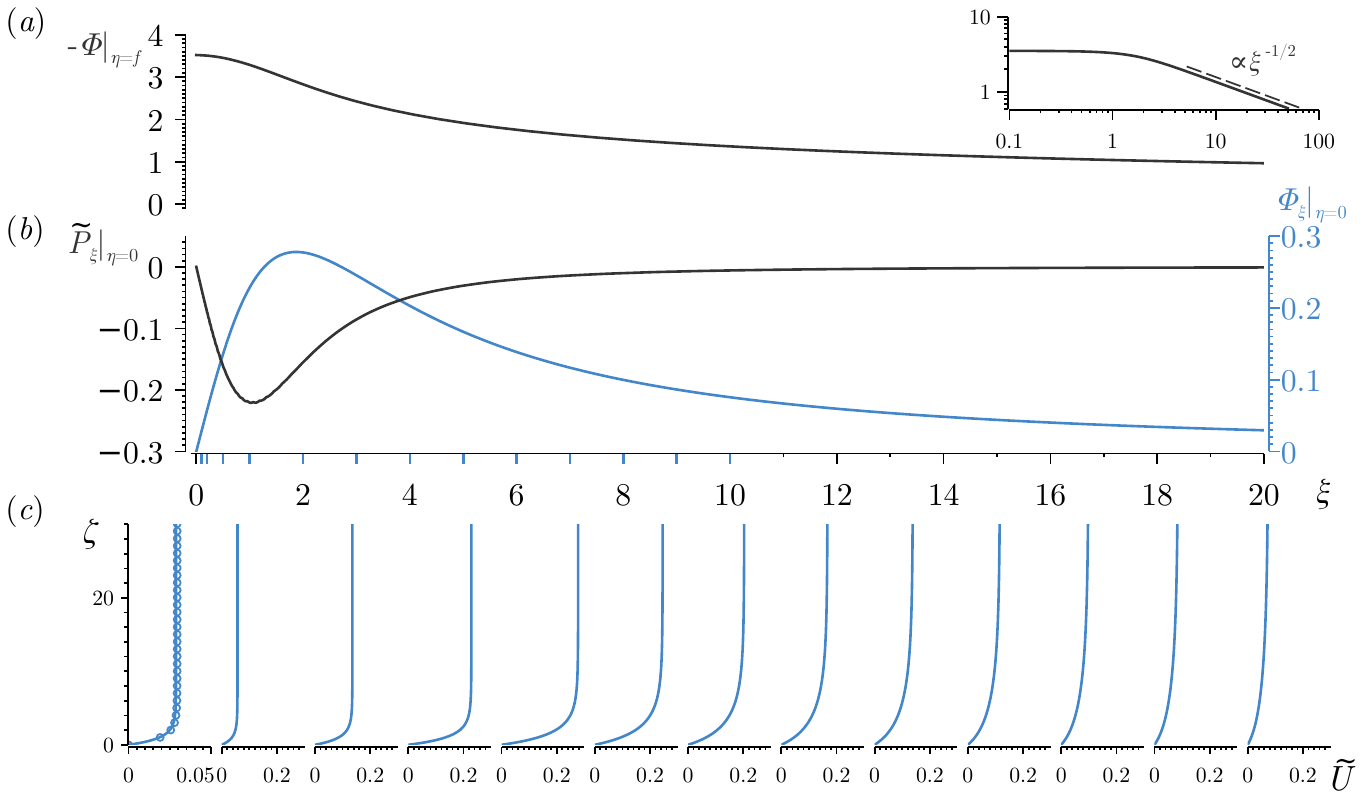}
    \caption{(a) Velocity potential at the free surface (solid line), and its asymptotic behaviour $-\Phi|_{\eta=f}\to \xi^{-1/2}$ for $\xi\gg 1$ (inset). (b) Wall slip velocity, $\Phi_{\xi}|_{\eta=0}$ (right axis), and wall pressure gradient, $P_{\xi}|_{\eta=0}$ (left axis). (c) Wall-boundary-layer velocity profiles at $\xi=(0.1,0.2,0.5,1,\ldots,10)$ (solid lines), and self-similar profile $\xi G'(\eta)$ at $\xi=0.1$ ($\circ$).}
    \label{fig:fig3}
\end{figure}

The potential solution described in~\S\ref{subsubsec:self} constitutes the outer flow for two viscous boundary layers sitting at the wall and at the free surface. In particular, the slipping velocity at the wall $\left. \Phi_\xi \right|_{\eta = 0}$, and the streamwise pressure gradient $\left.P_\xi\right|_{\eta = 0}$, evaluated numerically and presented in figure~\ref{fig:fig3}($b$), serve as boundary conditions for the wall boundary layer (WBL). The thickness of the WBL, $\delta_w^*$, is given by the balance $\rho u_c^{*2}/x_c^* \sim \mu u_c^*/\delta_w^{*2}\Rightarrow \delta_w^*\sim \mu\tau_*/\rho$, whence $\delta_w \sim \Oh^{1/2}\tau^{1/2}$. The wall shear stress (WSS), $\tauxy^*\sim \mu u_c^*/\delta_w^*\Rightarrow \tauxy^*\sim \mu^{1/2}\sigma^{11/8}/(\rho^{1/4}A^{5/8})\tau_*^{-11/10}$ which, so that the non-dimensional WSS writes $\tauxy \sim \Oh^{-1/2}\tau^{-11/10}$, as confirmed by the evolution of the shear stress in figure~\ref{fig:fig1}. The substitution into~\eqref{eq:ns} of the standard boundary-layer change of variables~\citep{rosenhead1963}, namely $ x = \tau^{2/5} \xi$, $y = (\Oh\tau)^{1/2} \zeta$, $u = \tau^{-3/5} \Ut $, $v = (\Oh/\tau)^{1/2} \Vt$, $p = \tau^{-6/5} \Pt$, and $h = (\Oh\tau)^{-1/2} \hht(\xi)$, yields the leading-order problem
\begin{subequations}\label{eq:BLmu}
\begin{alignat}{3}
    & \Ut_\xi + \Vt_\zeta = 0,  \label{eq:BLmucont}\\
    & \frac{3}{5} \Ut +   \left( \frac{2\xi}{5} + \Ut  \right)  \Ut_\xi + \left( \frac{\zeta}{2} + \Vt \right) \Ut_\zeta = -\Pt_\xi + \Ut_{\zeta\zeta}, \label{eq:BLmumom} \\
    & \Pt_\zeta = 0,
\end{alignat}
\end{subequations}
describing the WBL, so that the pressure gradient is given by the external potential flow as $-\Pt_\xi = \left. \left[(\Phi_\xi + 2 \xi /5)\Phi_{\xi\xi} + 3\Phi_{\xi}/5\right] \right|_{\eta=0}$. The boundary conditions accompanying the system~\eqref{eq:BLmu} are $\Ut=\Vt=0$ at $\zeta=0$, $\Ut\to \Phi_\xi$ as $\zeta \to \infty$, and $\Ut = 0$ at $\xi = 0$. The parabolic system~\eqref{eq:BLmu} is integrated using the method of lines by treating the streamwise coordinate $\xi$ as a pseudotime. Quadratic finite elements are employed in the transverse direction $\zeta$ and the discretised system is marched in $\xi$ using a second-order BDF algorithm with adaptive step size. To obtain the initial condition, advantage is taken of the self-similar nature of the boundary-layer for $\xi \ll 1$. Indeed, the substitution of a stream function $\xi G(\zeta)$ into~\eqref{eq:BLmucont}-\eqref{eq:BLmumom} yields,
\begin{equation}
    \label{eq:BLG}
    G'''+\left(G-\zeta/2\right)G''-G'\left(G'+1\right) = \xi^{-1}\Pt_\xi,
\end{equation} 
with the boundary conditions $G=G'=0$ at $\zeta =0$ and $G' \to \xi^{-1}\Phi_\xi$ as $\zeta \to \infty$. Equation~\eqref{eq:BLG} admits a self-similar solution for $\xi\ll 1$, in which $\xi^{-1}\Pt_{\xi}\to -0.3802$ and $\xi^{-1}\Phi_\xi \to 0.2965$, which was computed using a pseudospectral collocation technique~\citep{driscoll2014chebfun}.

Finally, for conciseness, let us just provide the characteristic scales of the free-surface boundary layer (FSBL), whose thickness $\delta_f^*$ can be estimated by taking into account that the free-surface vorticity is $\omega_f^*=-2\partial_s v_s^*+2u_s^*\kappa^*$ in a two-dimensional unsteady flow~\citep{lundgren1999}, where $\partial_s$ is the derivative along the interface, $u_s^*$ and $v_s^*$ are the tangential and normal velocity components at the interface, and $\kappa^*$ is the mean curvature. We now take into account that, in regime IV, $u_s^*\sim v_s^*\sim u_c^* \sim \hmind/\tau_*$ and $\partial_s \sim \kappa^* \sim 1/a$, providing $\omega_f^* \sim \hmind/(a\tau_*)$. Since by definition $\omega_f^* \sim \Delta u_s^*/\delta_f^*$, we deduce that $\delta_f^* \sim a \tau_* \Delta u_s^*/\hmind$, where $\Delta u_s^*$ is the characteristic velocity increment across the FSBL. The value of $\Delta u_s^*$ may be estimated by balancing the convective acceleration with the capillary pressure gradient, $\rho u_c^* \Delta u_s^*/\Delta x^* \sim \sigma \kappa^*/\Delta x^*\Rightarrow \Delta u_s^* \sim \sigma /(\rho a u_c^*)$. Thus, $\delta_f^* \sim \sigma/(\rho u_c^{*2}) \Rightarrow \delta_f^* \sim [\sigma/(\rho^{3/5}A^{2/5})]\,\tau_*^{6/5} \Rightarrow \delta_f \sim \tau^{6/5}$. We finally note that $\delta_f \to 0$ as $\tau \to 0$ much faster than the WBL, in agreement with the results shown in figure~\ref{fig:fig2}.

\section{Concluding remarks}

New insights about the inertia-driven dewetting of unstable ultrathin films of Newtonian liquids have been gained through theoretical analysis and numerical integration of the Navier-Stokes equations. We have shown that, when the Ohnesorge number $\Oh \lesssim 1$, the final approach of the flow towards the rupture singularity is self-similar, with a non-dimensional minimum film thickness $\hmin = 1.1\,\tau^{2/5}$ due to a dominant balance between liquid inertia and van der Waals forces. The spatial structure of the flow in the new regime presents a distinguished three-region structure characterised by a potential core separated by two viscous boundary layers, one sitting at the free surface and the other one adhered to the solid substrate. Making use of appropriate self-similar variables, the irrotational core has been described as a universal solution of the Euler equations, featuring a wedge-shaped interface with an opening angle of $45^{ \circ}$ off the solid. The latter non-slender solution applies when the minimum film thickness approaches the molecular scale, as similarly obtained by~\citet{morenobozaetal2020} in the context of viscous dewetting. A parameter-free description of the viscous wall layer, of non-dimensional thickness $\delta_w \sim \Oh^{1/2}\tau^{1/2}$, has been provided by integrating Prandtl's equations using self-similar boundary-layer variables. The free-surface vortical layer is much thinner than the viscous wall layer, since its thickness scales as $\delta_f \sim \tau^{6/5}$.

For finite values of the Onhnesorge number, an order-of-magnitude analysis allowed us to classify all possible regimes of dewetting, thereby unifying previous studies. The slender viscous regime discovered by~\citet{ZhangLister1999}, with $\hmin=0.913\,(\tau/\Oh)^{1/5}$, prevails for $\Oh\gtrsim 1$ and $\hmin(\tau) \gg 1$, provided that the initial thickness $h_o\gtrsim 30$. The non-slender viscous reported by~\citet{morenobozaetal2020}, where $\hmin=0.665\,(\tau/\Oh)^{1/3}$, applies for $\Oh\gtrsim 1$ and $\hmin(\tau) \lesssim 4$, independently of the value of $h_o$. The slender inertial regime described by~\citet{garg2017self} for shear-thinning fluids has been shown to apply also for Newtonian fluids, but only for $\Oh\lesssim 1$, $h_o\gtrsim 30$ and $\hmin(\tau)\gg 1$. The latter regime, where $\hmin=1.0\,\tau^{2/7}$, experiences a cross-over to the inertial regime discovered in the present work, which prevails for $\Oh\lesssim 1$ and $\hmin(\tau)\lesssim 4$ for any value of $h_o$.

Let us finally point out that, in contrast with polymer films, for which $\Oh\gg 1$, ultrathin liquefied metal films have associated values of $\Oh\lesssim 1$ or even $\Oh\ll 1$, as recently pointed out by~\citet{kondic2019liquid}. In such cases, the inertial regimes described in the present work may well be of practical relevance for the emergent field of liquid metal manipulation at sub-micrometer scales.

\begin{acknowledgments}
This paper is devoted to the memory of our beloved friend, Professor J. Fern\'andez-S\'aez (Pepe). His deep knowledge of mechanics, as well as his enthusiasm and kindness, have marked the lifes and careers of countless students and colleagues. Key numerical advice of Dr. J. Rivero-Rodr\'iguez is gratefully acknowledged. This research was funded by the Spanish MINECO, Subdirecci\'on General de Gesti\'on de Ayudas a la Investigaci\'on, through project DPI2015-71901-REDT, and by the Spanish MCIU-Agencia Estatal de Investigaci\'on through project DPI2017-88201-C3-3-R, partly financed through FEDER European funds. A.M.-C. also acknowledges support from the Spanish MECD through the grant FPU16/02562.
\end{acknowledgments}

\vspace{0.5cm}
\noindent\textbf{Declaration of Interests}\\
The authors report no conflict of interest.

%--------------------------------------BIBLIOGRAPHY---------------------------------------%

\end{document}